# Multitwist Möbius Strips and Twisted Ribbons in the Polarization of Paraxial Light Beams

Enrique J. Galvez[*], Ishir Dutta, Kory Beach, Jon J. Zeosky, Joshua A. Jones, and Behzad Khajavi

Department of Physics and Astronomy, Colgate University, Hamilton, New York 13346, USA

[*]Email: egalvez@colgate.edu

The polarization of light can exhibit unusual features when singular optical beams are involved. In 3-dimensional polarized random media the polarization orientation around singularities describe 1/2 or 3/2 Möbius strips. It has been predicted that if singular beams intersect non-collinearly in free space, the polarization ellipse rotates forming many-turn Möbius strips or twisted ribbons along closed loops around a central singularity. These polarization features are important because polarization is an aspect of light that mediate strong interactions with matter, with potential for new applications. We examined the non-collinear superposition of two unfocused paraxial light beams when one of them carried an optical vortex and the other one a uniform phase front, both in orthogonal states of circular polarization. It is known that these superpositions in 2-dimensions produce space-variant patterns of polarization. Relying on the symmetry of the problem, we extracted the 3-dimensional patterns from projective measurements, and confirmed the formation of many-turn Möbius strips or twisted ribbons when the topological charge of one of the component beams was odd or even, respectively. The measurements agree well with the modelings and confirmed that these types of patterns occur at macroscopic length scales and in ordinary superposition situations.

# INTRODUCTION

Electromagnetic waves can produce in 3-dimensions (3-d) unusual patterns of fields that are not immediately obvious. Superpositions of optical beams carrying scalar phase singularities, optical vortices, have been shown to produce knots in the line that follows the phase singularity in 3-d.[1-4] Vector fields can also produce unusual patterns. They may, for example, involve a traveling wave that when focused produces an axial component of the field. This is the case at the waist of a radially- or azimutally-polarized beam.[5] Because of this, they have long been discussed for use in charged-particle accelerators.[6] Radial and azimuthal vector beams are already a rarity in themselves because in their paraxial form they carry a polarization singularity: its center has all linear-polarization orientations. It is not a contradiction because, like with optical vortices, the intensity decreases asymptotically to zero at the singular point. These beams belong to a class of beams known as singular optical beams. Likewise, Poincaré beams can carry various types of polarization singularities. Among them, C-points, which have all orientations of the polarization ellipse, reached asymptotically from the polarization orientation of the surrounding field. The singular point in this case may have non-zero intensity because it has circular polarization, which is singular in orientation. The polarization field surrounding the C-point contains a disclination in the orientation of the polarization ellipse,[7-14] and recent studies have investigated the full range of disclinations that can be produced with designer singular optical beams.[15-17]

Evanescent fields are also situations that may give rise to unexpected arrangements of fields.[18,19] Freund proposed that electromagnetic fields involving randomly-polarized fields contain unique features: the polarization ellipse describes Möbius strips of either 1/2 or 3/2 turns along closed paths surrounding a singular point.[20-22] These features have been confirmed analytically by independent analysis.[23] However, they appear on length scales of the order of the wavelength, which makes them hard to measure.[22] Recently, Bauer et al. verified experimentally 1/2 and 3/2 Möbius strips appearing at the waist of a tightly-focused Poincaré beam.[24] They did so with the novel technique of Mie scattering nanointerferometry that they developed.[25] A more recent prediction claimed that singular optical beams would produce 3-d fields that describe many-turn Möbius strips at macroscopic length scales in the intersection of paraxial beams. These patterns would appear along a macroscopic closed path about the singular point.[26,27] Moreover, crossed paraxial beams with opposite circular polarization produce twists in the polarization along any closed paths with any type of modes.[26,27] The twists may be half-integer, forming Möbius strips only when the paths surround a polarization singularity.

The previous measurements of these types of features used a novel scattering method of tightly focused beams.[25] We made this determination using polarimetry. A particular challenge posed by this approach was that polarization projections cannot be used to measure the *z*-component of the field independently of the other components. However, we were able to derive it from projective measurements by relying on the simplified formalism of a symmetric geometric configuration. Using this method we were able to successfully reconstruct the 3-d orientations of the polarization ellipse, as shown below.

**MATERIALS AND METHODS**

**Theory**

The setup of the problem involves two paraxial beams of half-width $w$ crossing in free space. The beams were in orthogonal states of circular polarization. One beam was a standard Gaussian beam and the other carried an optical vortex, which for ease of analysis was prepared in a singly-ringed Laguerre-Gauss mode. This is depicted in Fig. 1(a). We analyzed the pattern in a reference frame ($x$, $y$, $z$), with $z$-axis coplanar with the propagation directions of the two beams, and bisecting them forming an angle $\theta$ with each direction, as shown in the figure. We label the $z = 0$ plane as the "observing plane." We can write the equations for the field at $t = 0$ in this plane as[28]

$$\vec{E} = (E_\ell + E_0)\cos\theta \, \hat{e}_x - (E_\ell - E_0)i \, \hat{e}_y + (E_\ell - E_0)\sin\theta \, \hat{e}_z \tag{1}$$

where

$$E_\ell = A_\ell \left(x^2 \cos^2\theta + y^2\right)^{|\ell|/2} e^{-(x^2\cos^2\theta + y^2)/w^2} e^{i[\ell\tan^{-1}(y/x\cos\theta) - kx\sin\theta]} \tag{2}$$

and

$$E_0 = A_0 e^{-(x^2\cos^2\theta + y^2)/w^2} e^{i(kx\sin\theta + \delta_0)} , \tag{3}$$

with $k$ being the wavenumber, $\ell$ the topological charge, $A_\ell$ and $A_0$ mode normalization constants, and $\delta_0$ the relative phase between the two beams. We neglect the curvature of the wavefront.

Because the field in Eq. (1) has all three components in a non-factorable form, the polarization-ellipse field is 3-dimensional. To qualitatively visualize the pattern that the ellipses make, let us assume that $\theta$ is a small angle, so that the $z$-component of the field is small. We express the $x$- and $y$-components of the field in terms of the right and left circular components: $\hat{e}_R = 2^{-1/2}(\hat{e}_x - i\hat{e}_y)$ and $\hat{e}_L = 2^{-1/2}(\hat{e}_x + i\hat{e}_y)$. Neglecting the z-component, the orientation of the polarization ellipse is half the relative phase between the circular components, or approximately

$$\varphi \approx \ell\phi/2 - kx\theta , \tag{4}$$

where $\phi$ is the transverse angular coordinate. For $\theta$ above $10^{-4}$ rad the variation in the $x$-dependence of the phase is larger than the azimuthal phase variation and the polarization displays polarization fringes, as shown in the measurements of Fig. 1(b), for which $\theta = 3 \cdot 10^{-4}$ rad.[28] In these fringes the polarization ellipse rotates by half a turn from fringe to fringe. So, if we follow a closed path around the center of the intersection of the two beams (the origin), the polarization ellipse rotates by $N$ half turns on the top side of the fringe pattern, and by $N+|\ell|$ half-turns in the opposite sense on the bottom side of the path. This leads to a net $|\ell|$ half-turns for a completed closed path.

Note also that in Eq. (1) the absolute value of the *y*- and *z*-components of the field are proportional to each other. Figure 1(c) shows the calculated amplitude squared of the *z*-component of the field. Adjacent fringes are out of phase with each other. That is, at the nodes of the fringe pattern the instantaneous *z*-component of the field changes sign, implying that the polarization ellipse is changing its tilt relative to the plane of the picture. In following this *z*-component along the circular path we see that it undergoes an odd number of tilt flips. When we combine the two-dimensional rotations with the *z*-component switching, we get that the semi-major axis of the polarization ellipse describes either a Möbius strip or a twisted ribbon.

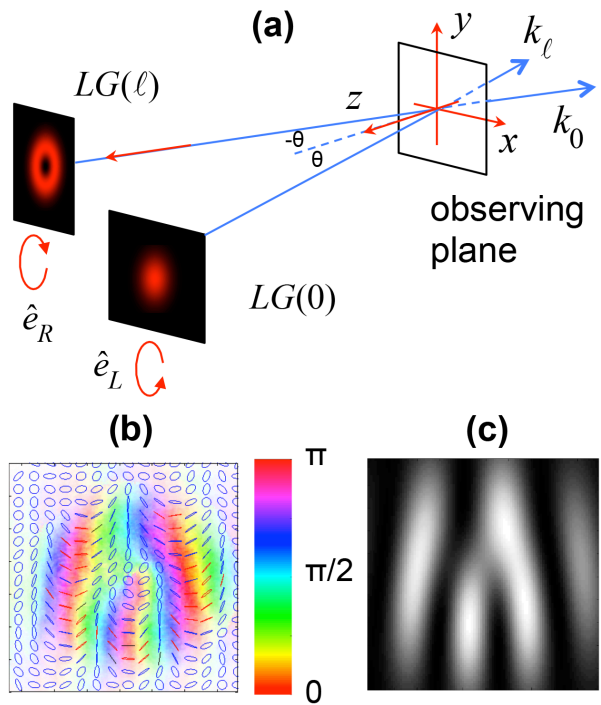

**Figure 1**. (a) Diagram of the situation that is studied: two beams in Laguerre-Gauss modes (LG) meet noncollinearly. One beam has right-circular polarization $\hat{e}_R$ and spatial mode with topological charge $\ell$, and the other has left-circular polarization $\hat{e}_L$ and fundamental Gaussian mode ($\ell = 0$). (b) 2-dimensional polarimetry of the light on the observing plane shows the colour-coded orientation of the polarization ellipse, and drawn ellipses coloured red/blue for right/left handedness. (c) Simulation of the amplitude squared of the *z*-component of the field.

We can quantify the twists the following way: the number of twists depends on the radius of the circular path $r$ relative to the fringe spacing $\lambda/(2\sin\theta)$,[27] where $\lambda$ is the wavelength of the light. The number of turns is then given by

$$N \simeq \frac{4r\sin\theta}{\lambda} + \frac{c}{2} \quad . \tag{5}$$

where $c=0$ for $|\ell|$ even, and $c=1$ for $|\ell|$ odd. Note that a twisted ribbon also appears for $\ell=0$, in the intersection of two plane waves. Indeed, the simple interference of non-collinear circularly polarized plane waves shows interesting features in its transverse momentum.[29]

**Projective Measurements**

The next problem we faced was the determination of the field of the light pattern. One approach that works with tightly-focused beams is to scatter the light with a small metallic sphere, and infer the field from the scattered pattern.[24,25] In our case the optical beams are unfocused, a situation where scattering may be too weak to use as means of measurement. We opted for using projective measurements with a polarizer. The challenge is that such measurements do not measure the $z$-component of the field independently of the other components.[30] However, as with polarimetry, one can obtain polarization-ellipse parameters via projective measurements plus knowledge of the modes of the polarization of the incoming beams and their geometric configuration. One additional restriction that simplifies the algebra is the use of an observing plane with a normal that bisects the two propagation-vector directions, as presented above in the setup of the problem.

To analyze the result of projective measurements we specify the field components of the two beams:

$$\vec{E}_\ell = E_\ell \left( \cos\theta \, \hat{e}_x - i\hat{e}_y + \sin\theta \, \hat{e}_z \right) \tag{6}$$

$$\vec{E}_0 = E_0 \left( \cos\theta \, \hat{e}_x + i\hat{e}_y - \sin\theta \, \hat{e}_z \right) \tag{7}$$

with respective propagation unit vectors:

$$\hat{k}_\ell = \sin\theta \, \hat{e}_x - \cos\theta \, \hat{e}_z \tag{8}$$

$$\hat{k}_0 = -\sin\theta \, \hat{e}_x - \cos\theta \, \hat{e}_z . \tag{9}$$

The intensity of the transmitted light past a polarizer with transmission axis along unit vector $\hat{p}$ is

$$I_p = \left| \left( \hat{e}_{p\ell} \cdot \vec{E}_\ell \right) \hat{e}_{p\ell} + \left( \hat{e}_{p0} \cdot \vec{E}_0 \right) \hat{e}_{p0} \right|^2 \tag{10}$$

where[30,31]

$$\hat{e}_{p\ell} = \frac{\hat{p} - \hat{k}_\ell(\hat{k}_\ell \cdot \hat{p})}{\sqrt{1 - (\hat{k}_\ell \cdot \hat{p})^2}} \tag{11}$$

is the unit vector orthogonal to the propagation direction along the plane that contains $\hat{p}$. We consider four polarization projections: $\hat{p}_V = \hat{e}_y$, $\hat{p}_H = \hat{e}_x$, $\hat{p}_D = (\hat{e}_x + \hat{e}_y)/\sqrt{2}$ and $\hat{p}_A = (\hat{e}_x - \hat{e}_y)/\sqrt{2}$, which correspond to vertical (*V*), horizontal (*H*), diagonal (+45 degrees; *D*), and anti-diagonal (−45 degrees; *A*) directions, respectively. The field component along the *y*-direction can be extracted directly from the projective measurement along *V*: $|E_y| = I_V^{1/2}$. However, we cannot separate the *x*- and *z*- components from the projections: $I_H = |E_x|^2 + |E_z|^2$. The field of Eq. (1) can be written as

$$\vec{E} = |E_\ell + E_0|e^{i\delta_+}\cos\theta\,\hat{e}_x + |E_\ell - E_0|e^{i(\delta_- - \pi/2)}\hat{e}_y + |E_\ell - E_0|e^{i\delta_-}\sin\theta\,\hat{e}_z, \tag{12}$$

which can be expressed to within an overall phase as

$$\vec{E} = |E_x|\hat{e}_x + |E_y|e^{i(\delta - \pi/2)}\hat{e}_y + |E_z|e^{i\delta}\hat{e}_z, \tag{13}$$

where $\delta = \delta_+ - \delta_-$. With the above equations we can establish relations between the amplitudes of the *z*- and *x*-components of the field, and extract them from the *V* and *H* projective measurements: $|E_z| = |E_y|\sin\theta = I_V^{1/2}\sin\theta$ and $|E_x| = (I_H - I_V \sin^2\theta)^{1/2}$. The phase $\delta$ is obtained from the *D* and *A* projections:

$$\sin\delta = \frac{(I_D - I_A)}{|E_x||E_y|}f(\theta), \tag{14}$$

where $f(\theta) = (1 + \cos^2\theta)^2 / (2\cos^4\theta - \cos^2\theta + 1)$. Thus, our measurement constitutes only a partial measurement of the fields because it uses *a priori* knowledge of the component beams to obtain the amplitudes of the *x*- and *z*-components of the field. We also used the fringe spacing from the data to obtain an accurate value of $\theta$. With this information we can reconstruct the field of Eq. (13). However, the expression only gives us an instantaneous value of the field. Our goal is to obtain the polarization ellipse. If we express the electric field in terms of the ellipse semi-axes, then it is given by

$$\vec{E} = \exp(-i\gamma)(E_a\hat{e}_a - iE_b\hat{e}_b), \tag{15}$$

where $\hat{e}_a$ and $\hat{e}_b$ are unit vectors along the semi-major and semi-minor axes of the ellipse, respectively, as shown in Fig. 2, and $E_a$ and $E_b$ are real coefficients. The phase $\gamma$ is related to the "rectification phase" $\beta = \tan^{-1}[(E_b/E_a)\tan\gamma]$, which is the angle that the instantaneous field vector makes with the semi-major axis of the ellipse,[9,32] as shown in Fig. 2. The phase $\gamma$ can be obtained directly from the field via the relation[33]

$$\exp(i\gamma) = \frac{\sqrt{\vec{E}\cdot\vec{E}}}{\left|\sqrt{\vec{E}\cdot\vec{E}}\right|}. \tag{16}$$

Thus, per Eq. (15), the semi-major and semi-minor axis vectors are respectively the real and imaginary components of the product of Eqs. (13) and (16).[27,33]

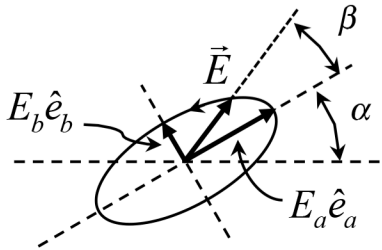

**Figure 2**. Polarization ellipse showing the relevant parameters: orientation $\alpha$, rectification phase $\beta$, and field amplitudes $E_a$ and $E_b$ along the semi-major and semi-minor axis directions $\hat{e}_a$ and $\hat{e}_b$, respectively.

**Apparatus**

A schematic of the apparatus is shown in Fig. 3. Light from a Helium-Neon laser was spatially filtered, expanded and sent through a Mach-Zehnder type interferometer where a spatial light modulator (SLM) was used as a folding mirror for both beams. Two panes of the SLM encoded spatial modes onto the light in first-order diffraction, as shown in the insert to Fig. 3. A polarizer insured that the polarization after the SLM was in a vertical plane. A half-wave plate in the path of one of the beams flipped the polarization to horizontal. The beams were recombined forming an angle $2\theta$ by a mirror and polarizing beam splitter. A quarter-wave plate converted the polarization states to the circular states. The beams overlapped at a digital camera closely preceded by neutral density filters and a film or wire-grid polarizer. The angle $\theta$ was obtained by measuring the fringe density of the interference pattern when both modes were fundamental Gaussian. We also did full polarimetric analysis of the light.

We took data for various angles, with the largest limited only by the camera resolution. We also took data for various topological charges of one of the beams. Images taken after the above mentioned 4 polarizer orientations were used to extract the semi-major axis of the polarization ellipse of every point in the image plane.

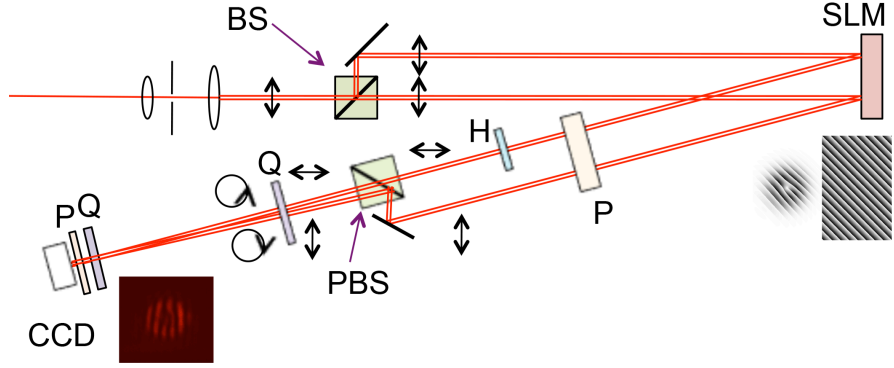

**Figure 3**. Schematic of the apparatus. Optical components include beam expander, spatial filter, non-polarizing beam splitter (BS), spatial light modulator (SLM), polarizer (P), half-wave plate (H), quarter-wave plate (Q), polarizing beam splitter (PBS) and digital camera (DC). Inserts show: a low resolution example of the patterns programmed onto the SLM, and image recorded by DC for a given polarizer orientation.

**RESULTS AND DISCUSSION**

Figures 4(a) and (b) show the modeled and measured fringe patterns for $\ell = 1$ at a shallow angle of $40 \pm 1$ arc sec. We also show in pane (c) the modeled 2-d view of the semi-major axis of the ellipse along a closed circle shown in pane (a), respectively, which describe a half-turn Möbius strip. Because we graph the semi-major axis vector ($E_a \hat{e}_a$), the evolution of the vector after a closed path yields the initial and final vectors parallel to each other but pointing in opposite directions. In pane (d) of the same figure we show the same quantity as extracted from the data. The colour-coding is labeled at the bottom of the figure. It helps in visualizing the 3-d orientation of the semi-major axis: red-magenta and blue/green are used when the vector describing the axis is above and below the observation plane, respectively. The red-green and magenta-blue are when the vector has a component that points toward or away the center of the circle, respectively.

The comparison of the two is in excellent qualitative agreement. This agreement is remarkable given the imperfections of the measured optical beams, which suffer aberrations, mode deformations due to the lack of mode purity and unwanted contributions from light for other diffraction orders. The adjustable parameters in the modelings were the relative phase between the two beams $\delta_0$, angle $\theta$, and the center of the pattern. The shortcoming is, of course, that we do not make independent measurement of the *x*- and *z*-components of the field and have to rely on knowledge of the geometry of the problem. The method could become more widely applicable to situations where the exact geometry is not known if a method to do independent measurements of the *x*- and *z*-components of the field is devised.

Note in Figs. 4(a) and (b) that the circle along which we do our calculation of the polarization ellipse barely goes over 1-2 fringes. In doing so, the Möbius strip carries only a half twist. As the circle covers more fringes, the polarization ellipse describes more twists.[27] We verified this as well, as shown below. The largest angle $\theta$ that we investigated was of about one degree, which involved about 7 camera pixels from fringe to fringe, yielding a Möbius strip with about 161.5 turns for a circular path with $r \simeq w$.

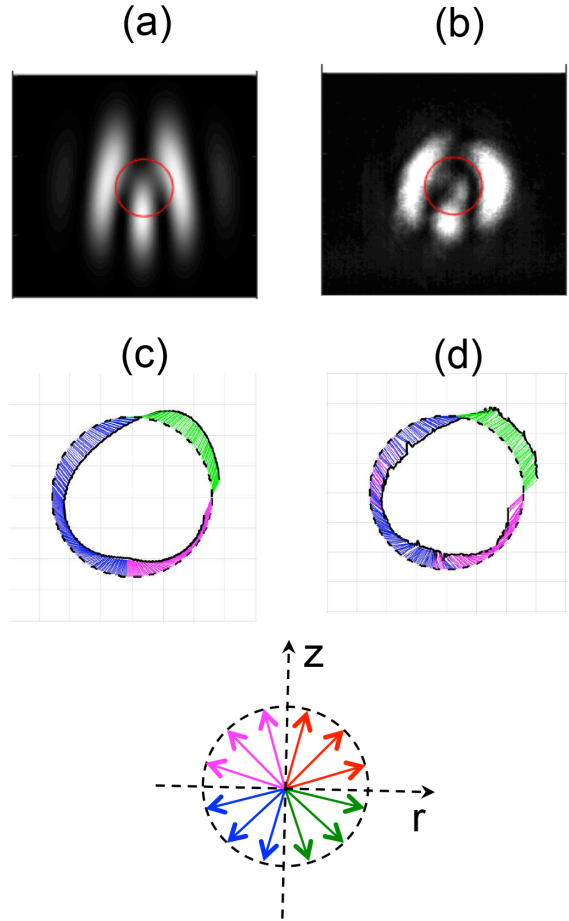

**Figure 4**. Modeled (a) and measured (b) projections of the light with a vertical polarizer. The circle denotes points along which we compute the semi-major axis of the ellipse of the modeled (c) and measured (d) data. Colour coding for the ellipse axis denotes the orientation in 3 dimensions: red-magenta and blue-green for above and below plane, respectively; red-green and magenta-blue with in-plane component pointing radially out or radially in, respectively. The coordinate $r$ lies within the $x$-$y$ plane with origin at the center of the circular path.

We experimented with other values of the topological charge, as shown in Fig. 5 for a few cases. It can be seen that for the same value of $\theta = 1.63 \pm 0.03$ arc min, the cases for $\ell = 1$ and $\ell = 3$ [(a), (b) and (e), (f), respectively] produced Möbius strips of 5/2 and 11/2 twists, respectively, whereas the $\ell = 2$ case [(c) and (d)] involves a ribbon of 3 twists. The 3-d views have different perspectives to better visualize the general character of the patterns. We added the vertical polarization projections to show the fringes that give rise to the modeled and measured reconstructions. We note that because the angle is shallow, the in-plane ($x,y$) coordinates are not in the same scale as the out-of-plane $z$-coordinate.

**Figure 5**. Modeled [(a), (c), (e)] and measured [(b), (d), (f)] 3-d orientations for $\theta = 1.63$ arc min but for

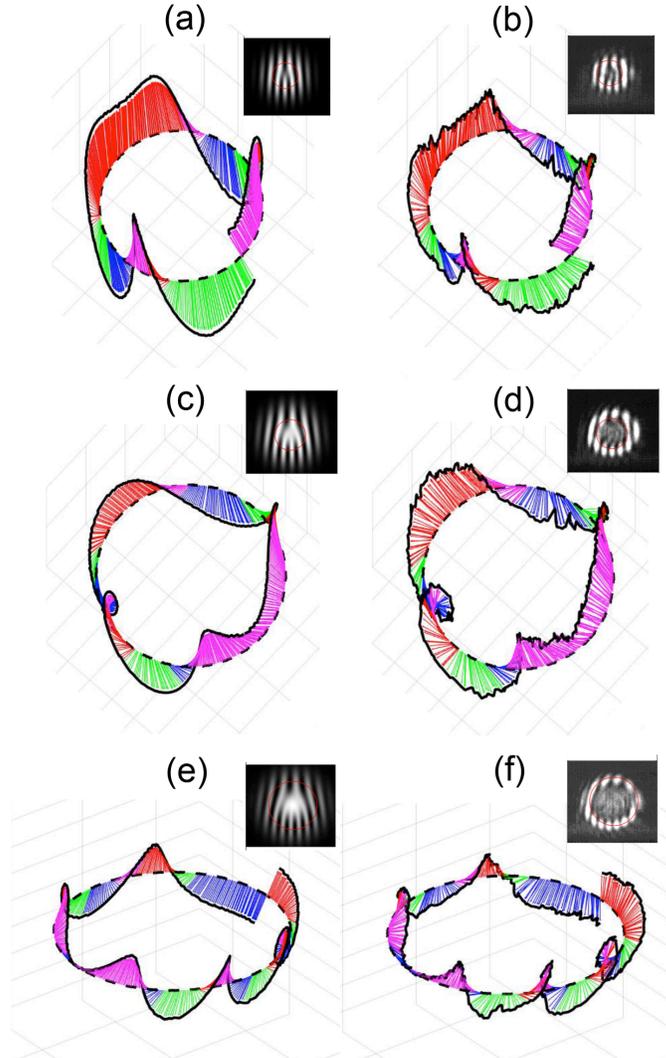

distinct values of topological charge: (a)-(b) for $\ell = 1$, (c)-(d) for $\ell = 2$, and (e)-(f) for $\ell = 3$. Inserts show the projective patterns in the vertical direction and the circular path corresponding to the reconstructions.

## CONCLUSIONS

In summary, we have confirmed the predicted patterns of twists in three dimensions that the polarization of the light describes when it is formed by the non-collinear interference of singular-optical beams. We confirmed that the polarization ellipse describes Möbius strips for $\ell$ odd and twisted ribbons for $\ell$ even.[27] Should both beams carry optical vortices of charge $\ell_1$ and $\ell_2$ the number of twists would be determined by the parity of $\ell = \ell_1 - \ell_2$ in Eq. (5).

We did these demonstrations at shallow angles to allow measurements to be done with a digital camera. The variations in the polarization are scalable via the angle formed by the propagation vectors of the two beams ($2\theta$). Patterns formed at larger angles, where all components are of comparable magnitude could not be imaged because the polarization fringe density was smaller than current camera pixel sizes. This work shows that non-collinear paraxial beams can be used to produce 3-d patterns that can be manipulated by adjusting the beam modes and their relative angle. If the medium were composed of molecules that interact strongly with polarization, these patterns could be used to manipulate them.[34,35] Beyond the fundamental interest, studies of 3-d polarization may have potential for adding polarization encoding to the storage of 3-d information. In this work we investigated the most basic situation: two paraxial beams with one of them bearing an optical vortex. The addition of more beams and modes is likely to show new interesting effects, opening a new dimension of complex light.

**ACKNOWLEDGMENTS**

This work was supported by the National Science Foundation grant PHY-1506321. We thank T. Bauer and W. Loeffler for useful discussions and S. Zhang for help.

**AUTHOR CONTRIBUTIONS**

EJG did the theory and analysis helped by ID; KB and JJZ took the data and JAJ and BK contributed to the analysis and conceptual understanding of the problem.